\documentclass{elsart}
 \usepackage{graphicx}
 \usepackage{epsfig}

\usepackage{amssymb}

\begin{document}

\begin{frontmatter}

\title{Proposing a new constraint for predictions of
$pp$, $\bar pp$ total cross sections and $\rho$ ratio at LHC}

\author{Keiji Igi$^a$ and Muneyuki Ishida$^b$}
\address{$^a$Theoretical Physics Laboratory, RIKEN, Wako, Saitama 351-0198, Japan}
\address{$^b$Department of Physics, School of Science and Engineering, Meisei University, Hino, Tokyo 191-8506, Japan}

\begin{abstract}
We propose a new constraint(1) corresponding to the FESR (with the moment n=-1) 
free from unphysical regions. Using this constraint(1) together with the constraint(2) 
(with the moment n=1), we search for the simultaneous best fit to the data points of 
$\sigma_{\rm tot}^{(+)}$ and $\rho^{(+)}$ ratio up to the SPS energies to determine 
those values at higher energies. 
We then predict $\sigma_{\rm tot}^{(+)}=107.1 \pm 2.6$ mb, $\rho^{(+)}=0.127 \pm 0.004$ 
at the LHC energy($\sqrt s$ =14 TeV). 
\end{abstract}

\end{frontmatter}


Recently\cite{[1],[2]}, we have searched for the simultaneous best fit of the average
of 
$pp$, $\bar pp$ total cross sections ($\sigma_{\rm tot}^{(+)}$),
and the ratio of the real to imaginary part of the forward scattering amplitude ($\rho^{(+)}$)
for 70 GeV $< P_{\rm lab} < P_{\rm SPS}$(up to the largest momentum of SPS 
corresponding to $\sqrt s$=0.9TeV) 
in terms of high-energy parameters constrained by the
finite-energy sum rule (FESR)\cite{[5]} with moment $n=1$.
We then predict $\sigma_{\rm tot}^{(+)}$ and $\rho^{(+)}$ in the LHC ($\sqrt s$=14TeV)
as well as high-energy cosmic-ray regions. Block and Halzen\cite{[3],[4]} have also rearched
the similar conclusions based on duality in a different approach.\\

\noindent \underline{\it Proposal of a new constraint}

The purpose of this Letter is to propose the other new constraint besides the previous one 
in order to constrain the above parameters.
Following ref.\cite{[1]}, we consider the crossing-even forward scattering amplitude defined by
\begin{eqnarray}
F^{(+)}(\nu ) &=& \frac{f^{\bar pp}(\nu )+f^{pp}(\nu )}{2}\ \ {\rm with}\ \ 
     {\rm Im}F^{(+)}(\nu )=\frac{k\sigma^{(+)}_{\rm tot}(\nu )}{4\pi}\ .
\label{eq1}
\end{eqnarray}
We also assume
\begin{eqnarray}
{\rm Im}F^{(+)}(\nu ) &=& {\rm Im}R(\nu )+ {\rm Im}F_{P^\prime}(\nu )\nonumber\\
 &=& \frac{\nu}{M^2}\left( c_0+ c_1 {\rm log}\frac{\nu }{M} 
     + c_2 {\rm log}^2\frac{\nu }{M}  \right)
    + \frac{\beta_{P^\prime}}{M} \left( \frac{\nu}{M}\right)^{\alpha_{P^\prime}}\ 
\label{eq2}
\end{eqnarray}
at high energies for $\nu > N$. 
Here, $M$ is the proton(anti-proton) mass and $\nu ,k$ are the incident
proton(anti-proton) energy, momentum in the laboratory system, respectively.
(We use the same notation as Ref.\cite{[1],[2]} in this article.)
Defining\footnote{
Although Re$F^{(+)}(\nu )$ becomes large for large values of $\nu$, a real constant 
has to be introduced in principle since the dispersion relation for Re$F^{(+)}(\nu )$
requires a single subtraction constant $F^{(+)}(0)$\cite{[6],[2]}.
}
\begin{eqnarray}
\tilde F^{(+)}(\nu ) &=& F^{(+)}(\nu ) - R(\nu ) - F_{P^\prime}(\nu) - F^{(+)}(0) 
      \sim \nu^{\alpha (0)}\ (\alpha (0) < 0)
\label{eq3}
\end{eqnarray}
for large value of $\nu$,
we have obtained\cite{[1]} in the spirit of $P^\prime$ sum rule\cite{[7]}
\begin{eqnarray}
{\rm Re}\tilde F^{(+)}(M) &=& 
\frac{2P}{\pi}\int_0^M \frac{\nu}{k^2}{\rm Im}F^{(+)}(\nu )d\nu 
+\frac{1}{2\pi^2} \int_0^{\overline{N}} \sigma^{(+)}_{\rm tot}(k)dk\nonumber\\
&&  - \frac{2P}{\pi} \int_0^N \frac{\nu}{k^2} 
      \left\{  {\rm Im}R(\nu )+\frac{\beta_{P^\prime}}{M}\left(\frac{\nu}{M}\right)^{0.5}
 \right\} d\nu\ ,
\label{eq4}
\end{eqnarray}
where $\overline{N}=\sqrt{N^2-M^2}\simeq N$. The equation (\ref{eq4}) was called 
FESR(1).\footnote{
This FESR(1) corresponds to $n=-1$\cite{[5]}.}
This FESR suffers from the 
unphysical regions coming from boson poles below the $\bar pp$ threshold.
Reliable estimates, however, are difficult.
Therefore, we have not adopted the FESR(1) in the analysis\cite{[1],[2]}.

Let us now change our strategy to use FESR(1), not at $\nu =M$ but
at some intermediate energy $\nu =\nu_1$ as 
\begin{eqnarray}
{\rm Re}\tilde F^{(+)}(\nu_1) &=& 
\frac{2P}{\pi}\int_0^M \frac{\nu {\rm Im}F^{(+)}(\nu )}{\nu^2-\nu_1^2} d\nu 
+\frac{P}{2\pi^2} \int_M^N \frac{\nu k}{\nu^2-\nu_1^2}\sigma^{(+)}_{\rm tot}(\nu )d\nu \nonumber\\
&&  - \frac{2P}{\pi} \int_0^N \frac{\nu}{\nu^2-\nu_1^2} 
      \left\{  {\rm Im}R(\nu )+\frac{\beta_{P^\prime}}{M}\left(\frac{\nu}{M}\right)^{0.5}
 \right\} d\nu\ .
\label{eq5}
\end{eqnarray}
If we choose the value of $\nu_1$ to be sufficiently large, this constraint is not sensitive
to the unphysical regions (the first term of the right-hand side of Eq.~(\ref{eq5})) as well as
ambiguities from low-energy integrals of $pp$ and $\bar pp$ scatterings.

Suppose we consider Eq.~(\ref{eq5}) with $N=N_1$ and $N=N_2$ ($N_2>N_1$). 
Taking the difference between these two relations, we obtain

\begin{eqnarray}
\frac{2P}{\pi}\int_{N_1}^{N_2} && \frac{\nu }{\nu^2-\nu_1^2} 
\left\{  {\rm Im}R(\nu )+\frac{\beta_{P^\prime}}{M}\left(\frac{\nu}{M}\right)^{0.5}
 \right\}  d\nu
 = \frac{P}{2\pi^2} \int_{N_1}^{N_2} \frac{\nu k}{\nu^2-\nu_1^2}
\sigma^{(+)}_{\rm tot}(\nu )d\nu \ \ .\ \ \ \ \ \ \ 
\label{eq6}
\end{eqnarray}
Let us call this relation as the \underline{constraint(1)} which we use in our analysis.
This constraint gives the relation between high-energy parameters $c_2,c_1,c_0,\beta_{P^\prime}$,
and the cross-section integrals, and is free from the unphysical regions.\\

\noindent \underline{\it The general approach}

Besides the constraint(1), we have the FESR corresponding to $n=1$\cite{[5]},
\begin{eqnarray}  &&
\int_0^M \nu Im\ F^{(+)}(\nu )d\nu 
     + \frac{1}{4\pi}\int_0^{\overline{N}} k^2 \sigma_{\rm tot}^{(+)}(k)dk \nonumber\\
 & &=  \int_0^N \nu Im\ R(\nu ) d\nu 
     + \int_0^N \nu Im\ F_{P^\prime}(\nu ) d\nu \ . \    
\label{eq7}
\end{eqnarray}
We call Eq.~(\ref{eq7}) as the \underline{constraint(2)} which we also use in our analysis.

The Re$F^{(+)}(\nu )$ is calculable from the Im$F^{(+)}(\nu )$, Eq.~(\ref{eq2})
by requiring the relation $F^{(+)}(-\nu )=(F^{(+)}(\nu ))^*$ to hold\cite{[8]}.
Therefore, we obtain \cite{[9]}
\begin{eqnarray}
\rho^{(+)}(\nu ) &=&  \frac{{\rm Re}F^{(+)}(\nu )}{{\rm Im}F^{(+)}(\nu )} \nonumber\\
  &=& \frac{   \frac{\pi\nu}{2M^2}\left( 
         c_1 + 2 c_2 {\rm log}\frac{\nu}{M} \right)  
        -\frac{\beta_{P^\prime}}{M} \left( 
         \frac{\nu}{M}\right)^{0.5}  +  F^{(+)}(0 ) }{
         \frac{k\sigma^{(+)}_{\rm tot}}{4\pi}
  } \ .
\label{eq8}
\end{eqnarray}

The constraints(1),(2) and the formula of $\sigma_{\rm tot}^{(+)}$(Eqs.~(\ref{eq1}) and (\ref{eq2}))
and the $\rho^{(+)}$ ratio (Eq.~(\ref{eq8})) are our starting points.
Among four parameters, $c_2,c_1,c_0$ and $\beta_{P^\prime}$, the $c_0$ and $\beta_{P^\prime}$
are represented by the other two ($c_2$ and $c_1$) parameters 
by two constraints. Then, we search for the simultaneous best fit to  
$\sigma_{\rm tot}^{(+)}$ and $\rho^{(+)}$ with three parameters, $c_2,c_1$ and $F^{(+)}(0)$.\\

\noindent \underline{\it Evaluation of cross-section integrals using experimental data}

The integrals of $\sigma_{\rm tot}^{(+)}$ appearing in RHS(LHS) of Eq.~(\ref{eq6})(Eq.~(\ref{eq7}))
are estimated by using experimental total cross sections $\sigma_{\rm tot}^{pp}$
and $\sigma_{\rm tot}^{\bar pp}$ in Particle Data Group 2004\cite{[10]}. 
A phenomenological formula, 
${\rm Im}f^{i}(\nu )(=\frac{k}{4\pi}\sigma_{\rm tot}^{i})$
$=\frac{\nu}{M}(c_0^i+c_1^i{\rm log}\frac{\nu}{M}+c_2^i{\rm log}^2\frac{\nu}{M})
+\frac{\beta^i}{M}(\frac{\nu}{M})^{0.5} +\frac{d^i}{M}(\frac{\nu}{M})^{-0.5} 
+\frac{f^i}{M}(\frac{\nu}{M})^{-1.5} $ for $i=\bar pp,\ pp$, is used to fit experimental  
$\sigma_{\rm tot}^{\bar pp}$($\sigma_{\rm tot}^{pp}$) in the region of 
2.5 GeV$\le k \le$100 GeV (2.592 GeV$\le k \le$100 GeV). The $c_2^i$ and $c_1^i$
are fixed with the values in our previous analysis, $(c_2^i,c_1^i)=(0.0479,-0.186)$
(``analysis 2" in ref.\cite{[2]}), and 
the 77(103) points of data
are fitted with four parameters, $c_0^i,\beta^i,d^i,f^i$, respectively. 
The best-fitted values of $\chi^2$ ($\chi^2/(N_D$-$N_P)$) are 148.5/(77-4) and 
71.8/(103-4), respectively for $\bar pp$ and $pp$. 
The large value of $\chi^2$ for $\bar pp$ comes from 
the inconsistency among the data of different experiments. 
In order to obtain good fit to $\bar pp$ data, we are forced to pick up 
some data points giving large $\chi^2$-contributions to be removed.
For this purpose we use statistical method, named Sieve algorithm\cite{Sieve,[3]}.
In this method, by minimizing the Lorentzian squared, 
$\Lambda_0^2=\sum_{i=1}^{N_D}{\rm ln}\{ 1+\gamma \Delta \chi^2_i \}$ 
(not the chi squared, $\chi^2=\sum_{i=1}^{N_D}\Delta \chi^2_i$)
with $\gamma =0.179$\cite{Sieve}, a ``robust" fit is obtained to the same data, 
where $N_D$ is the number of data points and $\Delta \chi^2_i$ means
the $\chi^2$-contribution of the $i$-th point. Points giving 
$\Delta \chi^2_i > \Delta \chi^2_{\rm max}$ in this robust fit are regarded as outliers, 
and removed. We take a cut $\Delta \chi^2_{\rm max}=6$, and seven points are 
removed.\footnote{
($k$(GeV),$\sigma_{\rm tot}^{\bar pp}$(mb))=(2.5,74.9$\pm$1.0), (3.54,69.7$\pm$0.5), 
(3.6,76.2$\pm$1.8), (4.,71.$\pm$ 1.), (4.015,66.84$\pm$ 0.32), 
(4.3,60.6$\pm$0.8), (9.14,57.51$\pm$0.73) are removed. 
}
After removing these points, we obtain a shifted data set, 
re-fitted by minimizing the conventional $\chi^2$.
As a result we obtain renormalized $\chi^2$(including the factor $R=1.140$\cite{Sieve}), 
$\chi^2_{\bar pp}/(N_D$-$N_P)$=$36.6/(70-4)$.
Finally we obtain the successful fits for both $\sigma_{\rm tot}^{\bar pp}$
and $\sigma_{\rm tot}^{pp}$. The 
$(c_0^i,\beta^i,d^i,f^i)=(6.34,11.08,5.28,-0.15)(\ (6.32,4.25,-12.64,24.4)\ )$ 
are obtained for $i=\bar pp(pp)$. 

We take the values of parameters appearing in Eq.~(\ref{eq6}) as 
$(\overline{N_1},\overline{N_2},k_1)=(10,70,40)$ GeV, where $\nu_1=\sqrt{k_1^2+M^2}$.
The above phenomenological fits give the cross-section integrals 
$\frac{P}{2\pi^2} \int_{\overline{N_1}}^{\overline{N_2}} \frac{\nu k}{\nu^2-\nu_1^2}
\sigma^{i}_{\rm tot}(k )dk = 242.57\pm 1.00\ (219.04\pm 0.47)$GeV$^{-1}$
with for $i=\bar pp(pp)$, where the errors correspond to the one-standard deviations. 
By averaging these values we obtain
\begin{eqnarray}
\frac{P}{2\pi^2} \int_{\overline{N_1}}^{\overline{N_2}} \frac{\nu k}{\nu^2-\nu_1^2}
\sigma^{(+)}_{\rm tot}(k)dk
 &=& 230.81 \pm 0.55\ {\rm GeV}^{-1}\ \ . \ \ \   
\label{eq9}
\end{eqnarray}
We can also evaluate the cross-section integral in Eq.~(\ref{eq7}).
We devide the region of integral into two parts,
$ \frac{1}{4\pi}\int_0^{\overline{N}} k^2 \sigma_{\rm tot}^{i}(k)dk
= \frac{1}{4\pi}\int_0^{\overline{N}^i_0} k^2 \sigma_{\rm tot}^{i}(k)dk
+ \frac{1}{4\pi}\int_{\overline{N}^i_0}^{\overline{N}} k^2 \sigma_{\rm tot}^{i}(k)dk$,
and the integral in higher energy-region(the second term) is evaluated by using the 
phenomenological fit in the same manner. 
The integral in lower energy region(the first term) is evaluated by using 
experimental data directly: Each datum is connected with the next point by a straight line in order,
and the resulting polygonal line graph gives the relevant integral.
(The details of this procedure are explained in our
previous works\cite{[1],[2]}.) By taking the $\overline{N}$ as 10 GeV and
$\overline{N}_0^{\bar pp}=4.7$GeV,
we obtain
$ \frac{1}{4\pi}\int_0^{\overline{N}} k^2 \sigma_{\rm tot}^{\bar pp}(k)dk
 = (522.22\pm 1.91)\ +\ (3499.44\pm 14.22) = 4021.66 \pm 14.35$GeV
for $i=\bar pp$.
By taking $\overline{N}_0^{pp}=4.966$GeV,
$ \frac{1}{4\pi}\int_0^{\overline{N}} k^2 \sigma_{\rm tot}^{pp}(k)dk
 = (357.24\pm 0.89)\ +\ (2411.26\pm 3.50) = 2768.50 \pm 3.61$GeV
for $i=pp$.
By averaging them we obtain
\begin{eqnarray}
\frac{1}{4\pi}\int_0^{\overline{N}} k^2 \sigma_{\rm tot}^{(+)}(k)dk
 = 3395.1\pm 7.4\ {\rm GeV}\ \ {\rm with}\ \  
         \overline{N}=10\ {\rm GeV},   
\label{eq10}
\end{eqnarray}
This value is consistent with our previous estimate, $3403\pm 20$GeV\cite{[1]},
which is evaluated by using the area of the polygonal line graphs up to $k$=$\overline{N}$(=10GeV).
In our present estimate,
both of the integrals are estimated with small errors less than 0.3\% .

\noindent \underline{\it FESR as two constraints}

By using the integrals, Eqs.~(\ref{eq9}) and (\ref{eq10}), we obtain the constraints(1) and (2) as
\begin{eqnarray} 
{\rm constraint(1)} && 3.316 \beta_{P^\prime}+ 31.98 c_0 + 141.1 c_1 + 610.9 c_2 = 230.81\ ,    
\ \ \ \ \ \ \label{eq11}\\
({\rm normalized}\ && 0.104 \beta_{P^\prime}+ c_0 + 4.41 c_1 + 19.1 c_2 = 7.22 \ ),
\nonumber\\
{\rm constraint(2)} && 140.7 \beta_{P^\prime}+ 383.6 c_0 + 781.6 c_1 + 1635. c_2 = 3395.1\ , 
\ \ \ \ \label{eq12}\\
({\rm normalized}\ && 0.367 \beta_{P^\prime}+ c_0 + 2.04 c_1 + 4.26 c_2 = 8.85 \ ),
\nonumber
\end{eqnarray}
where we neglect the errors of cross-section integrals, and regard Eqs.~(\ref{eq11}) and (\ref{eq12})
as exact constraints.
The equations are also rewritten in the form with the coefficient of $c_0$ normalized to unity 
in the parenthesis. 
Solving these two equations, we obtain the constraints for $c_0$ and $\beta_{P^\prime}$ as
\begin{eqnarray}
c_0=c_0(c_2,c_1) &=& 6.574-5.348 c_1 -24.95 c_2,\nonumber\\ 
\beta_{P^\prime}=\beta_{P^\prime}(c_2,c_1) &=& 6.206 + 9.025 c_1 + 56.40 c_2\ .\ \ \ 
\label{eq13}
\end{eqnarray}

\noindent \underline{\it Analysis and result} 

In the actual analysis we fit the data of Re$F^{(+)}(\nu )$ instead of $\rho^{(+)}$.
We made $\sigma_{\rm tot}^{(+)}$ and Re$F^{(+)}$ data points by averaging 
the original data given in ref.\cite{[10]}.
The detailed explanations for the treatment of data are given in ref.\cite{[1]}.
There are 17 data points of $\sigma_{\rm tot}^{(+)}$ above 70 GeV
and 10 data points of Re$F^{(+)}$ above 10 GeV
up to SPS energy $\sqrt s=0.9$ TeV.
They are fitted simultaneously with parameters $c_2,c_1$ for $\sigma_{\rm tot}^{(+)}$ and
$c_2,c_1,F^{(+)}(0)$ for Re$F^{(+)}$.
 
\begin{figure}
\includegraphics{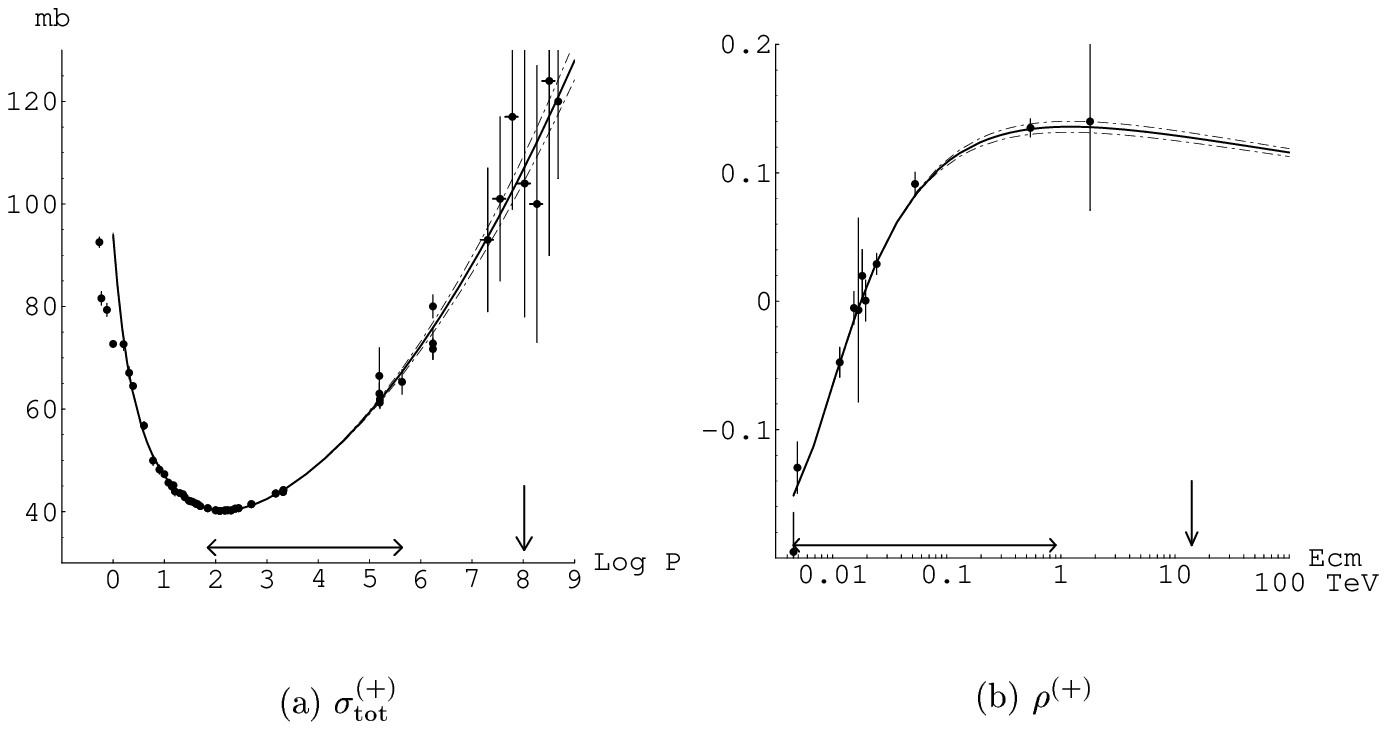}
\caption{\label{fig:1} Predictions for $\sigma^{(+)}$ and $\rho^{(+)}$:
(a) Total cross section $\sigma^{(+)}_{\rm tot}$} versus log$_{10}P_{lab}/$GeV,
(b) gives the $\rho^{(+)}(=Re\ F^{(+)}/Im\ F^{(+)})$ 
    versus $E_{cm}$ in terms of TeV. 
The fit is done for the data up to the SPS energy, in the region 
70(10)GeV$\le$ $k$ $\le$ 4.3$\times 10^5$GeV
(11.5(4.54)GeV $\le \sqrt s \le$ 0.9TeV) for $\sigma_{\rm tot}^{(+)}(\rho^{(+)})$ 
which is shown by horizontal arrow in each figure.
Vertical arrow represents the LHC energy $\sqrt s$=14TeV, corresponding to 
$k$=1.04$\times 10^8$GeV.
The thin dot-dashed lines correspond to the one standard deviation of $c_2$, given with 
the parameters $(c_2,c_1,c_0,\beta_{P^\prime},F^{(+)}(0))=
(0.0464\pm 0.0038,-0.158\mp 0.057,6.26\pm 0.21,7.40\mp 0.30,10.18\mp 0.27)$ .  
\end{figure}

The results are shown in Fig.~\ref{fig:1}.
The $\chi^2/d.o.f$ is $10.80/(27-3)$,which is less than unity.  
The respective $\chi^2$-values devided by the number of data points 
for $\sigma_{\rm tot}^{(+)}$ and $\rho^{(+)}$ 
are $\chi^2_\sigma/N_\sigma =5.64/17$ and $\chi^2_\rho /N_\rho =5.16/10$, respectively.
The fit is successful.

\begin{table}
\caption{
Values of parameters in the best fit using the constraint(1) with ($\overline{N_1},\overline{N_2},k_1$)
=(10,70,40)GeV and the constraint(2) with $\overline{N}$=10GeV.
The $c_0$ and $\beta_{P^\prime}$ are represented by $c_2$ and $c_1$ through the constraints, and the
fit is performed by using three parameters $c_2$, $c_1$ and $F^{(+)}(0)$, of which errors are given 
by the $\chi^2$ function $\chi^2(c_2,c_1,F^{(+)}(0))$.
Conversely by solving the constraints for $c_2$ and $c_1$,
the $\chi^2$ function is represented by $c_0$, $\beta_{P^\prime}$ and $F^{(+)}(0)$. 
The errors of $c_0$ and $\beta_{P^\prime}$ are, thus, obtained.
}
\begin{tabular}{ccccc|}
$c_2$ & $c_1$  & $c_0$  &  $\beta_{P^\prime}$ & $F^{(+)}(0)$\\
\hline
$0.0464\pm 0.0038$ & $-0.158\pm 0.057$ & $6.26\pm 0.21$ & $7.40\pm 0.31$ & 10.18$\pm$1.70 \\
\hline
\end{tabular}
\label{tab1}
\end{table}

The values of parameters are given in Table \ref{tab1}.
The result is compared with the previous analysis(``analysis 1" in ref.\cite{[2]}), 
where only Eq.~(\ref{eq12}) is used as a constraint.
The values of $c_2$ have to be noted
since the high-energy behaviours of 
$\sigma^{(+)}$ and $\rho^{(+)}$ are most sensitive to $c_2$.
The ``analysis 1" gives $c_2=0.0466\pm0.0047$.
The present value of $c_2$ in Table \ref{tab1} is consistent with the previous one, 
although the error does not reduce so largely. 

In order to check our result, we take another value of $k_1$, $k_1$=80GeV for constraint(1) with
the other parameters to remain unchanged ($(\overline{N_1},\overline{N_2},\overline{N})$=(10, 70, 10)GeV).
In this case the constraint(1) becomes
$-3.628 \beta_{P^\prime} -27.68 c_0 -113.0 c_1 -463.3 c_2 = -203.93\pm 0.34$GeV$^{-1}$
(normalized:$0.131 \beta_{P^\prime} + c_0 +4.08 c_1 + 16.7 c_2 = 7.37\pm 0.01$).
By using this together with the constraint(2)(Eq.~(\ref{eq12})),
the analysis is done in the same way. 
The value of $c_2$ is obtained as $c_2=0.0472\pm 0.0036$ with $\chi^2/d.o.f=10.84/(27-3)$.  
This $c_2$ is almost the same as in Table \ref{tab1}, and the result is considered to be 
almost independent of the value of $k_1$.

We have also checked the dependence of the errors of cross-section integrals,
Eqs.~(\ref{eq9}) and (\ref{eq10}).
In the case when a larger value of the integral, 230.81+0.55 (3395.1+7.4), is used 
for constraint(1) (constraint(2)) with the other integral to remain the same,
we obtain that the best-fit value of $c_2$ is 0.0485(0.0461) with $\chi^2=11.37(10.88)$.
The deviations of $c_2$ from the original value 0.0464 are 0.0021(-0.0003). 
They are small, compared with the statistical error 0.0038: about $(21^2$+$(-3)^2)/38^2$ = $30\%$.
So, we can regard Eqs.~(\ref{eq11}) and (\ref{eq12}) as exact constraints.

Special attention has to be paid when the two constraints
are employed to constrain the values of $c_0$ and $\beta_{P^\prime}$.
%
As shown in Eqs.~(\ref{eq11}) and (\ref{eq12}), the constraints(1) and (2) take the 
$c_0$-normalized forms
 $0.104\beta_{P^\prime}+c_0+\cdots$ and $0.367\beta_{P^\prime}+c_0+\cdots$, respectively.
They are linearly independent and we have obtained the meaningful result.
If the parameters are badly taken so that two constraints are not sufficiently linearly independent, 
the result becomes meaningless. For example, in case of $k_1$=60GeV, the constraint(1) has 
a normalized form $0.314\beta_{P^\prime}+c_0+\cdots$. It is quite close to the constraint(2),
and this selection of parameters are not suitable for the analysis. 
In the present analysis, in order to obtain sufficiently independent constraint, 
we have taken much larger value of $\overline{N_2}$(=70GeV) for constraint(1) 
than $\overline{N}$(=10GeV) for constraint(2). 

It is pointed out\cite{[4]} that there are strong resemblances between our approach 
and the one by Block and Halzen\cite{[3]}.
They estimated the values of experimental even-cross section $\sigma_{\rm even}(\nu_0)$
and of its derivative $\frac{d\sigma_{\rm even}}{d(\nu /M)}|_{\nu_0}$ at a certain energy 
$\nu =\nu_0=7.59{\rm GeV}(\sqrt s=4{\rm GeV})$ by using a local fit.
These two quantities are used as constraints for parameters 
$c_2$, $c_1$, $c_0$ and $\beta_{P^\prime}$, and $c_0$ and $\beta_{P^\prime}$ are 
represented by $c_2$ and $c_1$, similarly to our Eq.~(\ref{eq13}).
They have shown in Ref.[4] that the constraint for $\sigma_{\rm even}(\nu_0)$,
which gives $8.67=c_0+2.091c_1+4.371c_2+0.3516\beta_{P^\prime}$,\footnote{
In ref.\cite{[4]}, the constraint is given in unit of mb and the LHS
is given as 48.58 mb, which is replaced by 8.67 here in our notation where $c_i$'s are dimensionless.
}
is very close to our FESR(2), Eq.(12). 
Numerical difference seems very small at a first look,
but this difference is physically very important, since, in the former, the constraint
is obtained at one point $\nu =\nu_0$ in asymptotic region of $\sigma_{\rm tot}^{(+)}$,
while, in the latter, all the information in low-energy resonance region is included     
in the integral of $\sigma_{\rm tot}^{(+)}$ taken from $k =0$ to 10 GeV.

By using the values of parameters in Table~\ref{tab1},
we can predict the $\sigma_{\rm tot}^{(+)}$ and $\rho^{(+)}$  
 at Tevatron-collider energy($\sqrt s$=1.8TeV) and LHC energy($\sqrt s$=14TeV).
\begin{eqnarray}
\begin{array}{lcclcl}
\sigma_{\rm tot}^{(+)} &=& 75.82 \pm 1.02{\rm mb} & (\sqrt s=1.8{\rm TeV}), 
                        & 107.1\pm 2.6 {\rm mb} & (\sqrt s=14{\rm TeV}) \\  
\rho^{(+)}             &=& 0.136 \pm 0.004        & (\sqrt s=1.8{\rm TeV}),
                        & 0.127\pm 0.004        & (\sqrt s=14{\rm TeV})
\end{array}\ \ \ \ \ 
\label{eq14}
\end{eqnarray}
where
the relevant energies are very high, and
the $\sigma_{\rm tot}^{(+)}$ and $\rho^{(+)}$ can be regarded to be equal to the
$\sigma_{\rm tot}^{pp}$ and $\rho^{pp}$.

Our predicted values are almost the same as the previous ones\cite{[2]}.
They are consistent with the recent prediction by   
Block and Halzen\cite{[3]}
$\sigma_{\rm tot}^{pp}=75.19\pm 0.55$ mb, $\rho^{pp}=0.139\pm 0.001$ at Tevatron energy 
$\sqrt s=1.8$TeV, and 
$\sigma_{\rm tot}^{pp}=107.3\pm 1.2$ mb, $\rho^{pp}=0.132\pm 0.001$ at LHC energy 
$\sqrt s=14$TeV. They also analyzed the crossing-odd amplitude and 
obtained smaller errors compared with ours. 
Our prediction has also to be compared with 
Cudell et al.\cite{[11]} 
$\sigma_{\rm tot}^{pp}=111.5\pm 1.2_{\rm syst}\stackrel{+4.1}{\scriptstyle -2.1}_{\rm stat}$ mb, 
$\rho^{pp}=0.1361\pm 0.0015_{\rm syst}\stackrel{+0.0058}{\scriptstyle -0.0025}_{\rm stat}$,
whose fitting techniques favour the CDF point at $\sqrt s=1.8$TeV.

Finally we emphasize that our present analysis with two constraints is independent of 
the previous one\cite{[1],[2]} with one constraint. Although the high-energy parameters
are strongly constrained by two FESR, Eqs.~(\ref{eq11}) and (\ref{eq12}), in the present
analysis, the result is almost the same with the previous one\cite{[2]}.

It is worthwhile to point out the followings:\\
{\it 1}. Both of the parameters $c_0$, $\beta_{P^\prime}$ are constrained as
$c_0=c_0(c_2,c_1)$ and $\beta_{P^\prime}=\beta_{P^\prime}(c_2,c_1)$ through FESR (namely duality).\\
{\it 2}. And, then the high-energy behaviours of $\sigma_{\rm tot}^{(+)}$ have been $\chi^2$ fitted
in terms of the parameters $c_2$, $c_1$ since $\sigma_{\rm tot}^{(+)}$ is most sensitive to $c_2$.\\

\noindent{\it Acknowledgements} The authors express their gratitudes to professor M. M. Block
for carefully reading this paper and giving valuable comments to complete this work.



\end{document}